# Twist-Controlled Symmetry Breaking in Surface Phonon Polariton Moiré Metasurfaces

R. B. Iyer[1], S. H. Park[2,3], N. R. Sahoo[1], T. Low[2], and T. Folland [1*]

[1]Department of Physics and Astronomy, The University of Iowa, Iowa City, Iowa 52245 USA

[2]Department of Electrical and Computer Engineering, University of Minnesota, Minneapolis, Minnesota 55455, USA

[3]Department of Physics, The University of Texas at Austin, Austin, Texas, 78712, USA

**Abstract**

Moiré lattices provide a powerful route for engineering emergent symmetries and length scales through the relative rotation of periodic structures. However, their implementation in polaritonic systems remains relatively unexplored, and a general framework describing how twist modifies the interaction of optical modes in momentum space is still lacking. Here, we investigate how twist-induced moiré periodicities can control symmetry and momentum-space coupling in surface phonon polariton (SPhP) metasurfaces. We fabricate twisted overlapping dual-grating metasurfaces on a polar dielectric substrate with dielectric overlayer and characterize their optical response using polarization-resolved Fourier-transform infrared microscopy. Experimental measurements are combined with full-wave simulations and momentum-space analysis to identify the resonant SPhP and SPhP-like waveguide (WG) modes arising from both individual grating periodicities and emergent moiré periodicities. The results reveal twist-controlled symmetry breaking manifested as asymmetry between p→s and s→p polarization conversion, along with twist-dependent interactions between SPhP and SPhP-like WG modes. Our analysis reveals that the twist-engineered polarization-conversion asymmetry enables directional biasing of infrared radiative heat transfer. These findings establish twisted phonon-polaritonic metasurfaces as a versatile platform for geometry-controlled symmetry engineering in the mid-infrared. Future work may leverage such twist-programmable polaritonic interactions to enable directional thermal emission, polarization-selective detection, and reconfigurable infrared photonic devices.

## 1. Introduction

Moiré lattices have emerged as a powerful paradigm for realizing emergent length scales and modified symmetry properties that arise by superimposing periodic structures with a relative twist[1–5] or lattice mismatch[6,7]. For instance, in electronic systems, moiré lattices enable access to correlated and topological phenomena by restructuring band dispersions at longer periodicities. However, electronic moiré platforms are intrinsically constrained by material disorder, natural availability of crystal lattices, limited tunability of interlayer coupling, and the requirement of near-atomic precision in stacking and alignment. While electronic moiré systems emerge at atomic-scale periodicities (~1–10 nm), infrared photonics operates at much longer wavelengths

(hundreds of nanometers to microns), enabling precise, lithography-defined control over the emergent moiré length scale[8–15]. This motivates the exploration of moiré concepts in platforms governed by fundamentally different degrees of freedom and coupling mechanisms. Photonic structures provide a particularly appealing alternative in mid-infrared, offering direct access to momentum-selective interactions, broadband tunability, and design flexibility[11].

Within photonics, surface phonon polaritons (SPhPs) supported by polar dielectrics enable strong subwavelength confinement and low-loss propagation in the mid-infrared, making them attractive platforms for spectroscopy, thermal emission control, and infrared photonics[16,17]. Among SPhP materials, Silicon carbide (SiC) stands out for its broad Reststrahlen band spanning ~10.3–12.7 μm (800–970 cm$^{-1}$), providing ample low-loss spectral window for tightly confined polariton propagation[17–20]. This offers the spectral bandwidth and high-Q response necessary to engineer twist-dependent mode interaction, dispersion reshaping, and robust coupling effects essential for moiré metasurface designs in the mid-infrared. Instead of further increasing the patterning complexity of the polar substrate, we exploit a heterostructure approach, in which the intrinsic SPhP response of SiC is combined with an overlying photonic layer. In such SiC-based heterostructures, much of the metasurface functionality can be designed through engineering of the vertical stack, while leaving the polar dielectric largely unmodified.

Past demonstrations of moiré metasurfaces[11], have revealed interesting phenomena like magic-angle–like dispersion flattening[8,9,21,22], canalized polariton transport[21], and broadband polarization-selective optical activity controlled purely by twist angle or relative displacement[10,12,23,24]. Although recent studies highlight how infrared moiré metasurfaces can tailor nonlocal polaritonic responses and engineer effective symmetries at designer length scales, these advances are typically analyzed on a case-by-case basis. Additionally, in moiré systems, essential physics is most naturally described in momentum space[11,25], where multiple in-plane scattering channels coexist and interfere. The superposition of modes can give rise to symmetry breaking that is not rooted in local material anisotropy but instead emerges from geometry-dependent momentum-space selection[25–29]. Despite extensive studies of moiré phenomena in both electronic and photonic contexts, the specific role of the twist-generated periodicities in governing the momentum-space coupling and a framework for understanding twist-induced symmetry breaking in polaritonic photonic systems remains insufficiently clarified. Establishing a unified momentum-space framework is essential for evolving moiré concepts into scalable, symmetry-engineered platforms with broad relevance for infrared light–matter interaction.

Moreover, in contrast to conventional moiré systems based on vertically stacked layers, the moiré superlattice in our platform is realized entirely within a single planar metasurface. The twist is encoded through the relative rotation of two coplanar ribbon arrays patterned in the same metallic layer, so the usual notion of interlayer coupling or out-of-plane stacking does not apply. This purely planar implementation provides a fabrication-friendly route to moiré symmetry engineering while retaining precise control over the emergent momentum-space structure. Because the moiré pattern collapses onto a strictly planar structure, the metasurface remains achiral and does not support magnetoelectric responses. The optical behavior of such a structure can therefore be captured by an anisotropic electric surface conductivity[30].

## 2. SPhP Moiré Metasurface Design and Characterization Setup

In this work, we demonstrate twist-induced symmetry breaking in moiré phononic metasurfaces, revealing pronounced asymmetry in polarization conversion and reflectivity profiles arising from engineered SPhPs. Our design consists of a moiré-patterned metasurface formed by overlapping two grating-based metasurfaces with a controlled twist angle, as shown in figure 1a. This configuration induces (i) a reconfiguration of the azimuthal response that modifies rotational symmetry, yielding a directional geometrical asymmetry and (ii) tunable in-plane symmetry breaking through nonreciprocal polarization conversion, where s-to-p differs from p-to-s reflectivity at non-normal incidence due anisotropic surface conductivity. The structure consists of grating 1 (orange) and grating 2 (green), both possessing identical pitch and linewidth so that twist angle (α) serves as the sole source of structural asymmetry. When grating 2 is rotated relative to grating 1 by α, the interference between their periodicities gives rise to a moiré superlattice. In momentum space, the individual grating vectors $\mathbf{k}_{G0}$ and $\mathbf{k}_{G\alpha}$ combine vectorially to yield resultant momentums:

$$\mathbf{k}_{\text{Moiré-Long}} = \mathbf{k}_{G0} - \mathbf{k}_{G\alpha}$$

$$\mathbf{k}_{\text{Moiré-Short}} = \mathbf{k}_{G0} + \mathbf{k}_{G\alpha}$$

which defines the effective periodicity and in-plane symmetry of the composite metasurface. Here, we define the long-pitch moiré axis (LMA) and short-pitch moiré axis (SMA) as the in-plane directions associated with the emergent moiré reciprocal vectors corresponding to the longer- and shorter-effective moiré periods. We utilize the emergent moiré periodicities and their associated momentum channels to tailor and control the interaction between incident light, SPhPs in the underlying polar dielectric substrate, and SPhP-like waveguide (WG) modes supported by the dielectric overlayer. This momentum-dependent coupling impacts the optical response and gives rise to the observed twist-controlled symmetry breaking.

The principal optical modes excited within the designed metasurface under illumination is illustrated in figure 1b, using the simplified case of a single grating metasurface which was previously elucidated by Iyer et. al[31]. Incident light couples to SPhPs at the SiC interface (shown in blue along xz plane), with the grating providing the in-plane momentum necessary to bridge the wavevector mismatch. The amorphous silicon (a-Si) dielectric overlayer modifies the local dielectric environment, thereby tuning both the spectral position and spatial confinement of the SPhP resonance. As discussed earlier, in addition to SPhPs, a SPhP-like WG mode (depicted in purple on xz plane) is formed at the air/a-Si/SiC interface, exhibiting a pronounced $E_y$ field component and stronger confinement compared to a surface mode bound only by the air interface[32]In 1D gratings[31], these two modes can be treated as orthogonally oriented non-degenerate resonances, whose coupling governs polarization conversion. When their resonances are spectrally detuned and the excitation polarization is oriented at an intermediate angle, strong polarization conversion is equivalent to anisotropy in the in-plane effective permittivity tensor — a perspective that is particularly valid for 1D grating geometries where the conductivity tensor exhibits a clear principal-axis structure aligned with the grating periodicity. However, for the twisted moiré configuration studied here, this interpretation must be refined: the effective anisotropy is not simply a static permittivity tensor but rather arises from the momentum-dependent coupling of distinct twist dependent modes with different resonance frequencies and azimuthal response. By varying both the twist angle and the grating pitch, we achieve continuous control over the degree and nature of optical symmetry breaking, offering a versatile route to manipulate mid-infrared light–matter interactions. This approach can be extended to other SPhP materials and hence establishes a new paradigm for symmetry-engineered fabrication friendly photonic platforms, with promising implications for directional thermal emission, polarization-sensitive detection, and tunable infrared photonics.

To generate the angular and radial peak maps in Fig. 1 (c,d), we simulate real-space diffraction patterns of the twisted gratings over $\alpha = 0°–90°$, compute their 2D FFT power spectra, and extract peak power along angular ($\varphi$) and radial ($|k|$) directions. Here, $\varphi$ denotes the in-plane orientation angle of the incident electric field relative to the laboratory reference axis (LRA) defined by grating 1 ($\mathbf{k}_{G0}$). This analysis reveals how twist reorients the dominant momentum-space components and tunes the moiré length scales, providing a direct reciprocal-space visualization of the grating and moiré superlattices. The angular peak map in figure 1c shows how the dominant Fourier power in the azimuthal (angular) spectrum shifts linearly with twist angle, highlighting the well-defined angular branches associated with the underlying grating (0° and $\alpha$) and moiré symmetry directions. In contrast, the radial peak map in figure 1d tracks how the dominant spatial frequency of the diffraction pattern evolves with twist, revealing not only the branches associated with the original grating periodicities but also the emergence of two additional twist-dependent branches corresponding to the long-pitch and short-pitch moiré periodicities. In this radial representation, the lower-momentum branch and higher-momentum branch continuously tune with twist angle, directly visualizing the evolution of the long- and short-pitch moiré length scales in reciprocal space.

Figure 1e illustrates the conceptual design of the SPhP moiré metasurface and the corresponding optical measurement setup. Similar to structure in figure 1b, the moiré metasurface consists of a SiC substrate; however, in this case it supports two gold gratings, patterned in a single photolithography with identical pitch (varied between 2 µm and 9 µm) and fixed linewidth of 2 µm. A relative twist angle between the gratings (0°–90°) generates the moiré superlattice. An (a-Si overlayer deposited atop the gratings, serves both to tailor the local dielectric environment to enhance SPhP interactions and support SPhP-like WG modes. Optical characterization is performed using FTIR microscopy in a polarizer–analyzer configuration, allowing polarization-resolved reflectivity measurements. An aperture plate designed to mount over the microscope objective defines an annular illumination sector, enabling separation of p- and s-polarized components incident on the sample. When the input beam is polarized in the x–z plane, light emerging from the objective remains s-polarized in that plane and p-polarized in all others; rotation of the input polarization allows switching between s- and p-polarized excitation on the sample. This configuration is critical for accurately quantifying the in-plane birefringence of the moiré metasurfaces, as it ensures that the measured reflectivity arises solely from a well-defined polarization state. While annular illumination has been previously used in polarization microscopy[33], its implementation here at mid-infrared frequencies and with twist-controlled metasurfaces enables accurate and simultaneous quantification of polarization rotation and absorption. To our knowledge, such comprehensive polarization-resolved characterization has not previously been demonstrated in the mid-infrared. This technique is crucial for allowing unambiguous separation of symmetry-related contributions to the optical anisotropy. Measurements are carried out in both co-polarized and cross-polarized configurations to resolve polarization conversion effects. The spectra are acquired over 750–2000 cm$^{-1}$, encompassing the SiC Reststrahlen band and higher-order dielectric-guided modes. The sample is mounted on a motorized rotation stage, enabling precise control of its azimuthal orientation with respect to the incident beam. Reflectivity and polarization-conversion spectra are acquired as the sample is rotated from 0° to 360°, with all angles referenced to the axis of grating 1.

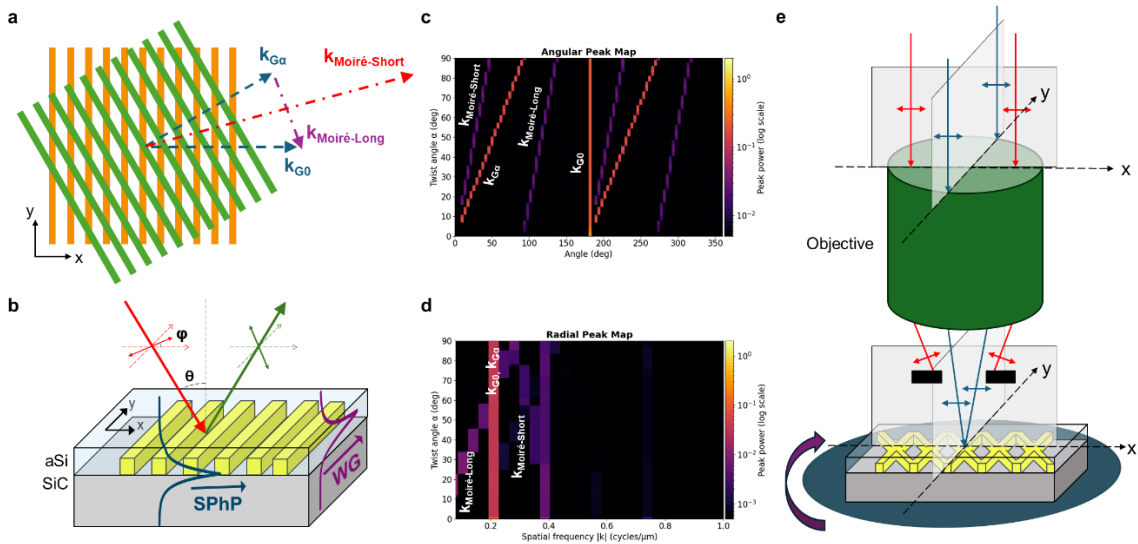

Figure 1: Conceptual overview of the moiré metasurface design and optical characterization. (a) Geometric construction of the twisted double-grating system, highlighting the formation of the Moiré grating vector (b) Illustration of key optical modes—SPhP and WG modes—excited under mid-IR illumination, (c) Angular peak map showing Fourier power shift with twist angle, (d) Radial peak map illustrating evolution of spatial frequencies with twist, and (e) Schematic of the sample structure and FTIR measurement setup.

## 3. Results and Discussions

Figure 2 presents polar color maps of co-polarized reflectivity for moiré metasurfaces with varying twist angles. The measurements span the spectral range of 750–1050 cm$^{-1}$, encompassing the SiC Reststrahlen band. Each map plots reflectivity as a function of azimuthal incident angle, defined with respect to the axis of the reference (orange in figure 1a) grating. Using this grating vector as the zero-angle direction defines the LRA. That is, the azimuthal angle ($\varphi$) is defined by the in-plane projection of the incident Poynting vector onto the metasurface, measured relative to this LRA. The top row corresponds to p-polarized excitation and the bottom row to s-polarized excitation, while the columns represent twist angles $\alpha = 30°$, $60°$, and $90°$. The s-polarized maps appear rotated by $90°$ relative to the p-polarized maps because switching from p to s input is implemented by rotating the incident electric field by $90°$ while keeping the aperture plate fixed and thus changes the way the light couples with metasurface. Since all azimuthal angles are still referenced to the LRA, this purely instrumental rotation produces an apparent $90°$ shift between the p- and s-polarized datasets, rather than reflecting an intrinsic symmetry difference of the metasurface. All samples share a grating pitch of 5.5 µm. Importantly, we show on the contour maps the LMA and SMA are rotated by $\alpha/2$ and $\alpha/2 - 90°$ respectively with respect to the LRA as consequence of the rotation of the electric filed.

Furthermore, the evolution of the co-polarized reflectivity with twist angle reveals systematic tuning of multiple resonant bands. In polar coordinates, the dominant spectral features appear as quasi-circular bands, whose behavior changes continuously with increasing twist. As $\alpha$ is increased from $30°$ to $60°$ and $90°$, these bands shift radially in frequency, undergo changes in curvature corresponding to variations in angular dispersion, and reorder in spectral proximity which is indicative of varying degrees of coupling between different resonant modes. This tuning behavior is observed consistently across all measured twist angles. At $\alpha = 90°$, the spectrum substantially simpler, with fewer dominant resonance dips exhibiting smoother and weaker angular dependence. The modes are largely confined below approximately 900 cm$^{-1}$, indicating diminished coupling efficiency between distinct in-plane momentum channels and a transition toward a weakly coupled regime, where distinct SPhP and waveguide branches remain largely non-interacting and the reflectivity shows only modest twist-induced angular dispersion. At $\alpha = 60°$, individual reflectivity dips become more clearly separated, indicating improved spectral distinction between modes. Angular modulation persists but is reduced for several resonances, consistent with a redistribution of oscillator strength between modes rather than the appearance or disappearance of resonances. At $\alpha = 30°$, the reflectivity spectrum is characterized by broad resonance dips whose angular tuning with $\varphi$ is complex to resolve due to their narrow spectral separation. Multiple bands coexist with comparable strength, resulting in significant spectral overlap and partially interacting resonances with weakly defined angular dispersion. However, modes in the 820–850 cm$^{-1}$ range tend to flatten near the LMA, while modes spanning 850–950 cm$^{-1}$ preferentially flatten near the SMA, suggesting that distinct momentum channels dominate different spectral regions. For reference, the untwisted ($\alpha = 0°$) grating metasurface exhibits a simplified spectral response, with all observed modes confined below approximately 900 cm$^{-1}$, providing a baseline that parallels the spectral compression seen at $\alpha = 90°$ (see S.I.1).

To further characterize the angular structure of the reflectivity beyond visual inspection, we performed full-wave finite-element simulations of the polarization-resolved response at varying twist angles, as described in S.I.2. Subsequently, we performed an angular Fourier decomposition of the azimuthal response of experimental data and simulated data, as detailed in S.I.3. This comparative analysis reveals a systematic redistribution of rotational harmonic content with twist angle, reflecting twist-controlled changes in the relative weights of different modes. Overall, these results show that twist serves as a continuous control parameter for engineering the coupling, mode interaction, and angular anisotropy of resonant modes in moiré metasurfaces, with maximal complexity in reflectivity spectra occurring at intermediate twist angles rather than at symmetry extremes.

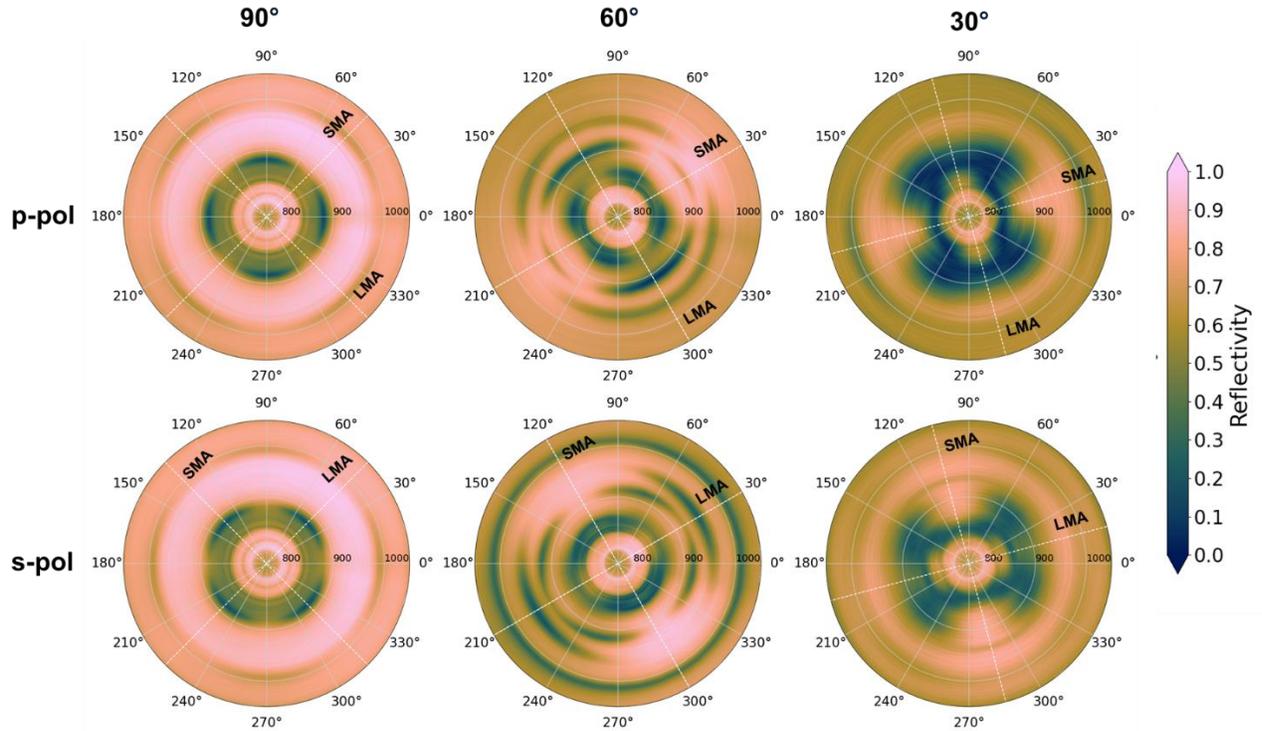

Figure 2: Polar color maps of co-polarized reflectivity for moiré metasurfaces with varying twist angles ($\alpha = 30°$, $60°$, $90°$). Each map shows reflectivity as a function of azimuthal angle and wavenumber (750–2000 cm$^{-1}$). Row 1: p-polarized input; Row 2: s-polarized input. Dotted lines indicate LMA and SMA

To reveal the continuous tuning of resonant modes controlled by the moiré superstructure, we probe the angular reflectivity data at varying grating twists angles along the LMA direction ($\varphi$ set to $\left(\frac{\alpha}{2} + \frac{\pi}{2}\right)$ relative to the LRA)—the direction of one of the resultant moiré grating vector (Fig. 1a). This novel analysis directly visualizes momentum-matching conditions for distinct SPhP/WG modes launched by grating vs. moiré periodicities as a function of relative rotation. Figure 3 presents experimental and simulated reflectivity color maps as a function of twist angle, with measurements and simulations performed for azimuthal incidence (electric field) aligned along the LMA direction. Multiple resonant modes appear in both datasets, showing rough qualitative agreement in their positions and dispersions, albeit with modest spectral shifts arising from the use of approximate dielectric optical constants adapted in simulations. Analytical waveguide mode dispersions, computed by using transfer-matrix theory, converted to equivalent twist angles, are overlaid on the simulation color map to validate the computational results against the analytical model (details in S.I. section 4). The calculations show that the modes near 850 cm$^{-1}$ correspond to SPhP modes launched by the moiré periodicity. Further, multiple additional highly dispersive modes emerge at higher wavenumbers in the Reststrahlen Band which have been identified as SPhP-like WG modes launched by the SPhP-like WG modes launched by the moiré periodicity. These WG modes exhibit modifications due to the input wavevector dependence, specifically through effective grating vectors given by $k_{WG}^M + k_0 sin(\theta_{inc})$ and $k_{WG}^M - k_0 sin(\theta_{inc})$, where the $k_{WG}^M$ is the moiré modified analytical waveguide-mode dispersion obtained from the transfer-matrix calculation (see S.I. Section 4) and incident angle projection $k_0 sin(\theta_{inc})$ tunes the momentum mismatch. This wavevector correction is necessary because the incident light provides an initial in-plane momentum component $k_0 sin(\theta_{inc})$, which must be accounted for in the reciprocal lattice vector summation to accurately phase-match the grating or moiré periodicity to the SPhP/WG dispersions in non-normal incidence geometries. We also note that $k_{WG}^M$ and $k_{WG}^M - sin(\theta_{inc})$ branches saturate beyond $\alpha \approx 35°$ since the mode cannot couple beyond the Reststrahlen band. Furthermore, the higher-frequency modes are attributed to higher-order waveguide resonances and higher-order grating diffraction components. By recasting the angular reflectivity data into twist-equivalent dispersion, these maps reveal the continuous tuning of modes and their interaction controlled by the moiré superstructure.

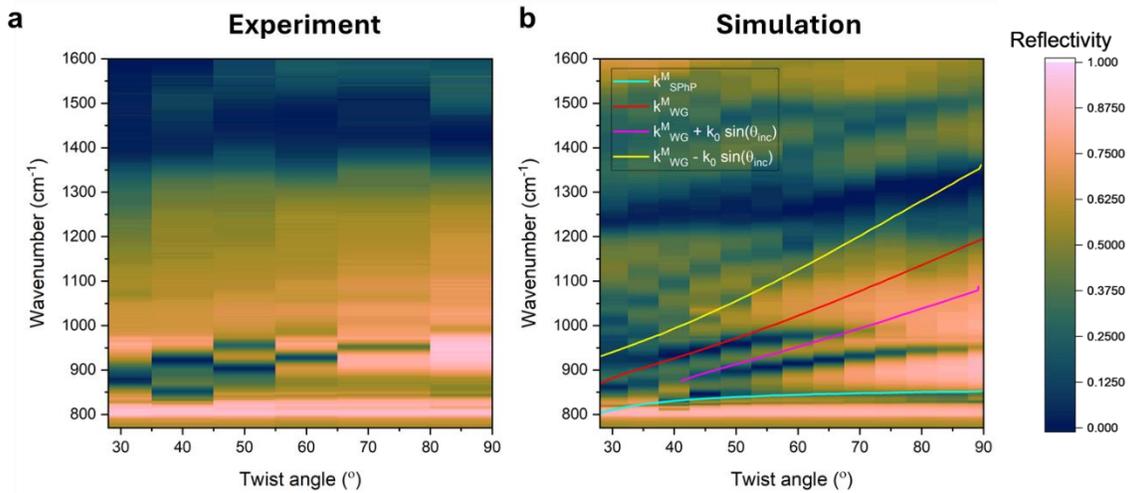

Figure 3. Experimental and simulated reflectivity color maps vs. twist angle equivalent (azimuthal incidence along LMA), showing SPhP and SPhP-like WG modes launched by grating and moiré periodicities, with analytical transfer-matrix dispersions overlaid on simulations

Figure 4 displays cross-polarized polar reflectivity color maps for the moiré metasurface, revealing polarization conversion effects and pronounced asymmetries in both conversion strength with respect to azimuthal angle and polarization response. The measurements were performed for a twist angle α = 30° and a grating pitch of 5.5 μm, with the two configurations corresponding to p → s and s → p conversion (previously noted in natural anisotropic materials[34] and complex metasurfaces[35,36]). The curvature dependence of modes is more prominently visible in the cross-polarized reflectivity data, strengthening the case for strong tunability of angular dispersion with azimuthal angle. Further, it is important to note that the p → s conversion channel exhibits higher efficiency than the s → p channel. This asymmetry between the p → s and s → p conversions can be captured by a polarization-resolved transfer-matrix model for a twisted anisotropic conducting sheet as discussed in detail in S.I. 6. The calculations in this framework explicitly demonstrate that $R_{sp} = |r_{sp}|^2$ and $R_{ps} = |r_{ps}|^2$ are different at non-normal incidence. Changing the twist angle shifts the modulation pattern of the cross-polarized reflectivity along the azimuthal angle and also modifies the overall magnitude of the conversion signals, reflecting the twist-dependent redistribution of coupling among momentum channels. But, the calculated $R_{ps}/R_{sp}$ ratio remains essentially independent of twist as shown in figure S7, indicating that the relative efficiency of p → s and s → p conversion is set primarily by the incidence geometry and anisotropic conductivity rather than by the moiré twist itself. However, the underlying anisotropic surface conductivity that sets this conversion asymmetry itself arises from the moiré geometry of the twisted double-grating structure.

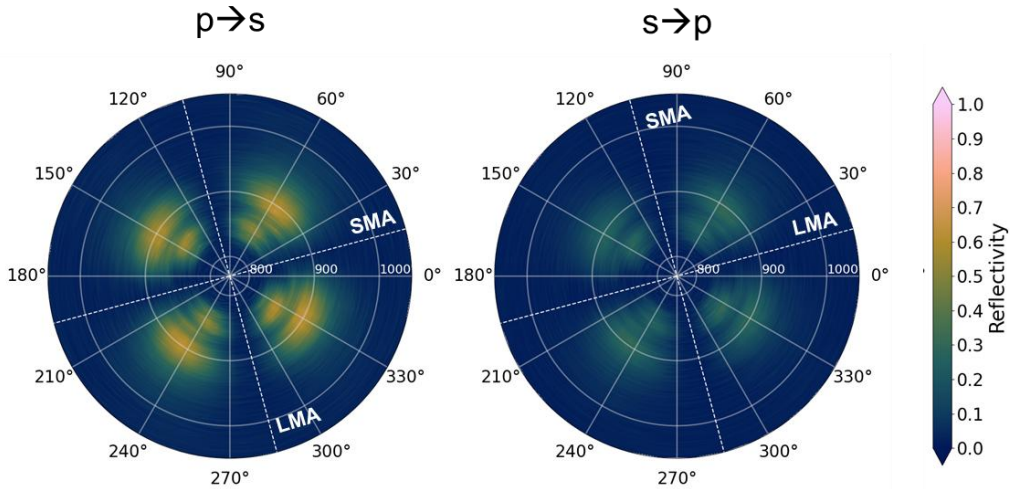

Figure 4. Cross-polarized reflectivity maps for the moiré metasurface at α = 30° and grating pitch of 5.5 μm, showing polarization conversion between p → s and s → p channels.

These results indicate that the moiré metasurface could serve as angularly selective "diodes" for mid-IR reflection. That is, by choosing an intermediate twist where the reflection asymmetry is strongest, the structure can preferentially reflect thermally generated radiation into one incident plane direction while suppressing in the opposite direction. Such twist-tunable, polarization-dependent angular control over reflectivity is directly relevant for thermal management and infrared camouflage systems that require enhanced reflection along specific escape channels while blocking others. To illustrate the thermodynamic implications of the polarization-conversion asymmetry, we consider a simplified radiative-exchange model in which two black bodies are coupled through the SPhP moiré metasurface (Fig. 5a). In this picture, body A emits predominantly p-polarized radiation while body B interacts with s-polarized radiation within the SiC Reststrahlen band. Because the metasurface converts polarization with unequal efficiencies $R_{p \to s}$ and $R_{s \to p}$, the radiative coupling between the two bodies becomes asymmetric. The band- and angle-limited radiative power emitted by each body into the illumination cone is obtained by integrating the Planck spectral radiance over the relevant spectral window and angular range. A detailed derivation of this model and the full expressions used in the calculations are provided in the S.I.section 6. Under these conditions, the net directional power bias mediated by the metasurface can be expressed as

$$P_{\text{net}} \propto \langle R_{p \to s} \rangle P_A - \langle R_{s \to p} \rangle P_B \quad (1)$$

where $P_A$ and $P_B$ are the band-limited emitted powers and $\langle R_{p \to s} \rangle$, $\langle R_{s \to p} \rangle$ denote the measured polarization-conversion efficiencies averaged over the spectral and angular ranges considered.

Figure 5b plots the normalized net directional power $\eta_{\text{net}} = P_{\text{net}}/P_{\text{BB}}$ for three representative temperature pairs, $(T_A, T_B) = (300,300)$ K, $(300,320)$ K, $and$ $(320,300)$ K using experimental polarization-conversion reflectivities averaged over the SiC Reststrahlen band. For equal temperatures, $\eta_{\text{net}}$ is non-zero solely due to the intrinsic polarization-conversion asymmetry $R_{p \to s} \neq R_{s \to p}$. Its magnitude decreases with increasing twist angle, consistent with the weakening of cross-polarized reflectivity contrast at larger twist. When $T_A > T_B$, the overall $\eta_{\text{net}}$ increases at all twists, whereas for $T_B > T_A$ the overall $\eta_{\text{net}}$ decreases. We emphasize that this calculation is a proof-of-concept based on band-averaged reflectivities over the SiC Reststrahlen band and a simplified angular cone, intended to illustrate the utility of twist-engineered polarization conversion for biasing radiative heat flow. Further design optimization can substantially enhance these directional efficiencies for practical thermal management applications. Overall, these trends confirm that the twisted moiré metasurface acts as a tunable, polarization-selective thermal mediator whose directional efficiency depends both on twist-controlled mode coupling and on the imposed temperature gradient between the radiatively coupled black bodies.

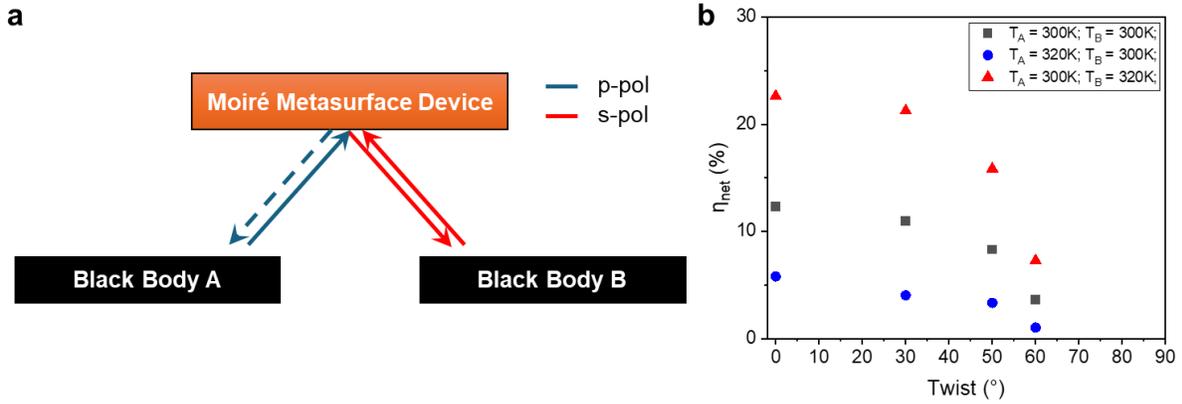

Figure 5. Proof-of-concept model for twist-induced asymmetric radiative heat exchange. (a) Schematic for polarization-mediated radiative exchange through a moiré metasurface device and (b) net radiative bias efficiencies calculated from experimentally measured reflectivity

## 4. Conclusion

In summary, we have introduced twist-controlled SPhP moiré metasurfaces on SiC substrates with a-Si overlayers, enabling precise engineering of SPhPs and SPhP-like waveguide modes through emergent moiré periodicities formed by twisted gold gratings. Polarization-resolved FTIR measurements and finite-element simulations revealed twist-tunable symmetry breaking, including nonreciprocal p→s and s→p polarization conversion asymmetry, azimuthal reconfiguration of reflectivity dips, and mode interactions. The modes launched by moiré periodicity and additional individual grating branches were validated by analytical transfer-matrix theory. These results demonstrate fabrication-friendly control over mid-IR momentum-space symmetries without substrate etching. Furthermore, our analysis demonstrates that this twist-engineered asymmetry enables directional biasing of infrared radiative heat transfer—even at $\Delta T=0$ K—where the metasurface preferentially transmits p→s converted radiation from one blackbody while suppressing the reciprocal s→p channel, establishing a purely geometric thermal diode without temperature gradients or active components. Beyond SiC substrates and a-Si overlayers, these concepts and techniques can be extended to diverse polar dielectric materials across frequency regimes, establishing our design as a versatile platform for geometry-controlled polaritonic symmetry engineering. Thus, our study advances beyond case-by-case moiré photonic studies by providing a unified framework for geometry-induced anisotropy in polaritonic systems. Furthermore, this platform unlocks applications in directional thermal emission, IR camouflage, and polarization-sensitive detection, paving the way toward scalable, twist-reconfigurable metasurfaces across SPhP materials for active mid-IR photonics.

**Acknowledgements**

T.G.F. and R.B.I. gratefully acknowledge support from the University of Iowa College of Liberal Arts and Sciences and the Department of Physics and Astronomy at the University of Iowa. Financial support from the Office of Naval Research through the Multi-University Research Initiative (MURI) on Twist-Optics (Grant No. N00014-23-1-2567) is acknowledged by T.G.F., R.B.I., and T.L. Portions of this research were performed using facilities at the Materials Analysis, Testing, and Fabrication (MATFab) Facility. Additional experimental work was carried out at the Minnesota Nano Center (MNC), which is supported by the National Science Foundation through the National Nanotechnology Coordinated Infrastructure (NNCI).